\documentclass{aa}  

\usepackage{lscape}
\usepackage{graphicx}
\usepackage{txfonts}
\begin{document}

   \title{A phylogenetic approach to chemical tagging}

   \subtitle{Reassembling open cluster stars}

   \author{Sergi Blanco-Cuaresma
          \inst{1,2}
          \and
          Didier Fraix-Burnet
          \inst{3}
          %\fnmsep\thanks{}
          }

   \institute{Observatoire de Gen\`eve, Universit\'e de Gen\`eve, CH-1290 Versoix, Switzerland
   \and
   Harvard-Smithsonian Center for Astrophysics, 60 Garden Street, Cambridge, MA 02138, USA \\
   \email{sblancocuaresma@cfa.harvard.edu} 
\and 
Univ. Grenoble Alpes, CNRS, IPAG, Grenoble, France\\
              \email{didier.fraix-burnet@univ-grenoble-alpes.fr}
             %\thanks{}
             }

   \date{Received May 28, 2018; accepted July 10, 2018}

% \abstract{}{}{}{}{} 
% 5 {} token are mandatory
 
  \abstract
  % context heading (optional)
  % {} leave it empty if necessary  
  {The chemical tagging technique is a promising approach to reconstruct the history of the Galaxy by only using stellar chemical abundances. Different studies have undertaken this analysis and they raised several challenges.}
% aims heading (mandatory)
{Using a sample of open clusters stars, we wish to address two issues:
minimize chemical abundance differences which origin is linked to the evolutionary stage of the stars and not their original composition; evaluate a phylogenetic approach to group stars based on their chemical composition.}
 % methods heading (mandatory)
{We derived differential chemical abundances for 207 stars (belonging to 34 open clusters) using the Sun as reference star (classical approach) and a dwarf plus a giant star from the open cluster M67 as reference (new approach). These abundances were then used to perform two phylogenetic analyses, cladistics (Maximum Parsimony) and Neighbour-Joining, together with a partitioning unsupervised classification analysis with k-means. The resulting groupings were finally confronted to the true open cluster memberships of the stars. } 
% results heading (mandatory)
   {We successfully reconstruct most of the original open clusters when carefully selecting a subset of the abundances derived differentially with respect to M67. We find a set of eight chemical elements that yields the best result, and discuss the possible reasons for them to be good tracers of the history of the Galaxy.}
  % conclusions heading (optional), leave it empty if necessary 
   {Our study shows that unraveling the history of the Galaxy by only using stellar chemical abundances is greatly improved provided that i) we perform a differential spectroscopic analysis with respect to an open cluster instead of the Sun, ii) select the chemical elements that are good tracers of the history of the Galaxy, and iii) use tools that are adapted to detect evolutionary tracks such as phylogenetic approaches.}


   \maketitle
%
%-------------------------------------------------------------------

%\listoftodos

\section{Introduction}

Stars are known to typically form in large number within molecular clouds which generally have a homogeneous chemical composition \citep{2014Natur.513..523F}. The newly born stars may rest gravitationally bound for very long periods of time (e.g., the open clusters that we observe today), but most of the stellar aggregates will dissolve due to gravitational interactions. The stars may end up with very different kinematics and their original chemical composition is the main clue that can help us identify family of stars born together or, at least, at similar periods of time \citep{2002ARA&A..40..487F}.

To distinguish and cluster the different families by individually tagging stars, we need to consider a rich set of individual chemical abundances derived homogeneously and apply clustering techniques. Such multivariate clustering (unsupervised classification) approaches were already performed by several studies such as \citet{Blanco-Cuaresma2015} using the k-means technique applied on several Principal Component Analysis components, \citet{Hogg2016} with a k-means technique on 15 abundances, and \citet{Smiljanic2017} with a hierarchical clustering technique. 

The resulting clustering showed how challenging the problem is even with very homogeneous spectroscopic analysis and high quality observations. The chemical composition of the stars is altered with time due to stellar evolution (e.g., atomic diffusion), which at the same time is different depending on the original stellar mass. The simplifications and assumptions that our current models incorporate also affect the results in different ways depending on the stellar evolutionary stage (e.g., NLTE effects). At the very least, these effects need to be minimized to improve the results generated by classification algorithms. But, additionally, the clustering techniques used by previous studies may not be the most suitable for solving the problem.

The most important caveat of these partitioning (such as k-means) and hierarchical techniques is that they are based on similarities, they are thus adapted to detect distinct and more or less compact structures in the multivariate parameter (here abundance) space. This could be a valid approach if all the surface chemical abundances (i.e., the abundances we can derive from spectroscopy) do not change with time, in that case each open cluster would always define a "cloud" of stars in parameter space defined by the chemical abundances. But this is not the case since stars of the same family evolve differently according to their mass, leading each family of stars to be distributed along elongated structures that partitioning and hierarchical techniques are not designed for.

In addition, partitioning techniques require the number of groups to be input a priori. In other words we have to guess the number of families in advance, which is impossible in practice when applying the chemical tagging technique to field stars. There are a number of statistical criteria that can be used for this purpose, but this is often not very conclusive as it will be briefly discussed in this paper.

The multivariate methods used so far are thus not well adapted to be used in the context of the chemical tagging technique. However, it has been shown that phylogenetic approaches are able to reconstruct the evolutionary tracks of stars from abundances \citep{Fraix-BurnetHouches2016}. At first glance, stellar evolution does not appear to be a good case for a phylogenetic study since there seems to be no direct transmission of properties between stars. But, firstly, the evolution of a star depends on mainly two parameters (their initial mass and metallicity), identical stars at different ages are related together, their relationships being represented by the evolutionary track. Secondly,  the enriched gas expelled through the explosions or winds of the most massive stars form the new generation of stars. For instance, this is the basis of the well-known Populations I, II and III in galaxies. This is a typical transmission with modification process between different families of stars, that can be representing as a branching pattern linking the different evolutionary tracks.

For evolutionary tracks, the transformation driver is the age of the stars, so that each family of stars is characterized by the same initial set of chemical abundances and mass. While when trying to identify stars born together during the same period of time via chemical tagging, the transformation driver within each open cluster is mass and not age. To avoid confusion with the traditional evolutionary tracks of stars most often shown on the HR diagram, we will call "family tracks" the paths followed by stars born together. Such family tracks are exactly the equivalent of the ones used to determine the age of stellar populations in globular and open clusters (i.e. isochrones) except that we are here in a $p$-dimension space (the $p$ abundances).

A first phylogenetic analysis of real stars in our Galaxy was performed by \citet{Jofre2017}. Their goal was not to classify stars following the chemical tagging objectives, that is to group stars that were born together, but to obtain an evolutionary scenario for stars in the solar neighborhood. They used a Neighbour Joining technique, which like cladistics (Maximum Parsimony), is a phylogenetic tools seeking to establish relationships between the objects under study (see for instance \citet{Fraix-Burnet2015} and Appendix~\ref{AppendixMethods}). From the relationships depicted by their resulting tree, they draw conclusions about the star formation rate in the thick and thin disc, as well as investigate the impact of processes such as radial migration and disc heating. This confirms the usefulness of the phylogenetic approaches for stellar astrophysics.

In the present paper, we use the relationships to reconstruct the different families but we do not seek to discuss the relationships between them. We want to test the ability of phylogenetic approaches to group stars applying the chemical tagging technique on a sample of open cluster stars. One key element of our study is that phylogenetic reconstructions rely on "characters" which are descriptors (observables, parameters, variables or properties), such as chemical abundances in our case, acting as tracers of the history of the Galaxy (see Appendix~\ref{AppendixMethods}). As a consequence, it is often counter-productive to use all available observables blindly, and we show in this paper that limiting the analysis to a selection of certain chemical abundances leads to a much better result.

This paper is organized as follows. The data are presented in Sect.~\ref{data} and the methods in Sect.~\ref{method}. The cladistics (Maximum Parsimony) results are shown in Sect.~\ref{results}, together with the results from Nieghbor Joining and k-means. We discuss some possible explanation for the selected set of abundances Sect.~\ref{discussion} and conclude in Sect.~\ref{conclusion}. Appendix~\ref{AppendixMethods} presents a detailed description of the methods used and  Appendix~\ref{TabContingencies} contains the contingency tables for the four biggest open clusters.

%--------------------------------------------------------------------
\section{Data}
\label{data}

To obtain high-precision chemical abundances it is necessary to use high-resolution stellar spectra with a good signal-to-noise ratio (S/N). The dataset presented in \cite{Blanco-Cuaresma2015} is very convenient for this work, it contains 2'121 spectra of FGK stars observed with three different instruments:

\begin{itemize}
\item NARVAL: $\sim$300 to 1100~nm and an average resolution of $\sim$80'000.
\item HARPS: $\sim$378 to 691~nm with a gap between chips that affects the region from 530 to 533~nm, and a resolution of $\sim$115'000.
\item UVES: $\sim$476 to 683 nm with a small gap between 580 and 582~nm, and a minimal resolution of  $\sim$47'000.
\end{itemize}

The observed spectra were re-analysed using iSpec\footnote{\url{http://www.blancocuaresma.com/s/}} \citep{Blanco-Cuaresma2014}, a spectroscopic tool that has been significantly improved since the dataset was first published (e.g., spectral normalization can now be done using synthetic templates).

After co-adding all the spectra corresponding to the same star and observed with the same instrument/setup, the dataset is composed of 446 stellar spectra. Considering the mean radial velocity of each cluster \citep{Blanco-Cuaresma2015}, we discarded the stars with radial velocity 2.5~km/s greater/lower than the reference one or with errors higher than 1.5~km/s. The final dataset is composed of 371 stellar spectra.

Following the method described in \cite{Blanco-Cuaresma2016}, we derived atmospheric parameters (i.e., effective temperature, surface gravity, metallicity) and individual absolute chemical abundances for all the stars and the Sun. Then, two different sets of line-by-line differential abundances are computed using different reference stars:

\begin{itemize}
\item With respect to the Sun (discarding stars with surface gravities between $3.00$ and $4.00$ dex to be coherent with the M67 set).
\item With respect to the open cluster M67 using the star No164 for giants (i.e., $log(g) < 3.00$ dex) and No1194 for dwarfs (i.e., $log(g) > 4.00$ dex).
\end{itemize}

For certain stars, we had more than one observation made with different instruments (i.e., they could not be co-added) and we selected the one with higher S/N. This process leave us with a dataset of chemical abundances for 207 stars when compared to the Sun, and 180 stars when compared to \object{M67}. The former corresponds to 34 open clusters, while the latter to 33 as shown in Table~\ref{TabClusters}.

From this analysis we cannot obtain the exact same number of chemical elements for all the stars, there are targets and reference stars that do not show reliable absorption lines for certain elements. Hence, we measure up to 29 chemical abundances in total but we cannot derive all these abundances for all the stars in the dataset. We could limit the analysis to the minimum common set, but the classification technique explored in this study can deal with a reasonably low number of unknown values.

Here after, when we refer to a chemical element symbol we mean its chemical abundance relative to the iron abundance of the same star (e.g., references to Ti represent [Ti/Fe]), except for iron, which is relative to hydrogen (i.e., [Fe/H]). All of them averaged from the differentially line-by-line calculation with respect to a reference star (the Sun or M67 depending on the dataset as explained above). The differential individual chemical abundances computed here for each star are used as input variables for the methods explained in Sect.~\ref{method}, the average values and dispersion per open cluster when using M67 as reference are summarized in Tables~\ref{tab:abundances_summary_msel}, \ref{tab:abundances_summary_rest1} and \ref{tab:abundances_summary_rest2}. and discussed in Sect.~\ref{discussion}.

    \begin{table}
       \caption[]{Cluster names, number of stars per cluster, and index of cluster used on some figures.}
          \label{TabClusters}
\begin{tabular}{lrrcc}
             \hline
             \noalign{\smallskip}
\hfil Cluster & \multicolumn{2}{c}{Number of stars}& \multicolumn{2}{c}{Index of cluster}  \\            
\hfil Name   &    wrt Sun \hfil &     wrt M67 \hfil  & Sfull & Mfull / Msel \\
             \noalign{\smallskip}
             \hline
             \noalign{\smallskip}
 M67          &    35  &   23     & 1 & 1       \\
 NGC6705      &   20   &   20     & 2 &  2  \\
 IC4651       &    20  &   19     & 3 &  3    \\
 NGC2632      &   19   &   19     & 4 &  4  \\
 IC2714       &     8  &    8     & 5 &  5      \\
 NGC2539      &     8  &    8     & 7 &  6    \\
 NGC752       &     7  &    7     & 8 &  7     \\
 Melotte111   &     6  &    6     & 9 &  8    \\
 NGC2360      &     8  &    6     & 6 &  9    \\
 NGC2447      &     6  &    6     & 10 &  10   \\
 NGC4349      &     6  &    6     & 11 &  11    \\
 NGC3680      &     5  &    5     & 12 &  12    \\
 NGC1817      &     4  &    4     & 16 &  13    \\
 NGC2567      &     4  &    4     & 17 &  14    \\
 NGC6494      &     4  &    4     & 18 &  15    \\
 NGC6940      &     5  &    4     & 14 &  16    \\
 Melotte22    &     3  &    3     & 20 &  17      \\
 NGC3114      &     3  &    3     & 22 &  18    \\
 NGC3532      &     3  &    3     & 23 &  19    \\
 NGC6633      &     4  &    3     & 19 &  20   \\
 IC4756       &     4  &    2     & 15 &  21      \\
 Melotte71    &     2  &    2     & 24 &  22     \\
 NGC2099      &     2  &    2     & 25 &  23    \\
 NGC2251      &     2  &    2     & 26 &  24    \\
 NGC2287      &     2  &    2     & 27 &  25    \\
 NGC2423      &     3  &    2     & 21 &  26    \\
 NGC2548      &     2  &    2     & 28 &  27     \\
 NGC5822      &     5  &    2     & 13 &  28    \\
 Collinder350 &     1  &    1     & 30 &  29      \\
 Melotte20    &     1  &    1     & 31 &  30      \\
 NGC2477      &     1  &    1     & 32 &  31     \\
 NGC2547      &     1  &    1     & 33 &  32     \\
 NGC6475      &     1  &    1     & 34 &  33     \\
 NGC6811      &     2  &   --     & 29 & --        \\
              \noalign{\smallskip}
             \hline
             \noalign{\smallskip}
Total & 208 & 183 & & \\
   \noalign{\smallskip}
             \hline
          \end{tabular}
    \end{table}

%--------------------------------------------------- Two column table
    \begin{table*}
       \caption[]{Groups of selected chemical elements tested with the different methods presented in this work}
          \label{TabAbundances}
\begin{tabular}{l l}
             \hline
             \noalign{\smallskip}
             Sample analyzed     & Chemical elements \\
             \noalign{\smallskip}
             \hline
             \noalign{\smallskip}
      Sfull &      Al Ba C  Ca Ce Co Cr Cu Fe La Mg Mn Na Nd Ni S  Sc Si Sm Sr Ti V  Y  Zn Zr \\
    Mfull  & Al Ba C  Ca Ce Co Cr Cu Eu Fe La Mg Mn Mo Na Nd Ni O  Pr S  Sc Si Sm Sr Ti V  Y  Zn Zr  \\
  Msel & Al Ba Co Fe Mg Mn Sc Ti  \\
  \noalign{\smallskip}
             \hline
          \end{tabular}
    \end{table*}

\section{Method}
\label{method}

\subsection{Phylogenetic and Partitioning Analyses}

Phylogenetic techniques look for the relationships between objects (stars in this work), minimizing a path through all the objects of the sample, providing a graph representation of the simplest scenario for transforming any object into any other one by changing their descriptors (individual chemical abundances in this case). The result of a phylogenetic analysis is thus a tree. Each family of stars is supposedly spread along a track which appears as a sub-structure (bunches of branches) on the tree. 

Cladistics, also called Maximum Parsimony (hereafter MP), is the simplest and most general phylogenetic technique: among all the possible tree arrangements of the sample objects, we select the one which possess the least number of changes on all its branches. These changes are counted from the discretized input variables (i.e., chemical abundances) with a simple L1-norm (Manhattan distance). This approach has been explained in detail in many papers for astrophysics \citep[e.g. ][]{jc1,jc2,Fraix-Burnet2015,Fraix-BurnetHouches2016} and is summarized in Appendix~\ref{AppendixMethodsMP}.

The MP analyses were performed with the software PAUP 4.0a154\footnote{\url{http://paup.phylosolutions.com/}} \citep{paup}.  The quality of the family reconstruction was estimated by eye with the help of the representation shown on Fig.~\ref{FigTrees}. Due to the hierarchical nature of the phylogeny, there is some arbitrariness when defining the families. We decided to remain conservative by selecting the largest structures that the tree suggests without going too much into subdivisions (see gray boxes in Fig.~\ref{FigTrees}), placing ourselves in a real life situation where we would know nothing about the families. The tree structures chosen and shown in Fig.~\ref{FigTrees} were established independently of the real cluster membership and correspond to visually obvious groups of branches. Naturally, this grouping can be especially disputed for the many small clusters (i.e., composed by very few stars) for which any global quantitative measures of the similarity between two data clusterings such as the Adjusted Rand Index is not informative (i.e., it would be equally possible to choose different break-points in the graph and make different boxes). This is not the case for the four biggest open clusters M67, NGC6705, IC4651 and NGC2632 for which we compute quantitative estimates of our results.

The trees resulting from MP analyses are unrooted, i.e. there is no evolutionary direction, unless some common ancestorship is chosen. The analysis does not depend on this choice, so that rooting a tree only modifies its graphical representation and therefore may somewhat impact the groupings from its structure and the associated interpretation. For convenience in this paper, we have rooted all the trees with the stars having the lowest values for logg.

For comparison and reliability estimation, we have also performed a phylogenetic analysis with a different technique, the Neighbor Joining Tree Estimation algorithm,  and a partitioning analysis with the well-known k-means method using the function \textit{kmeans} of the package \textit{stats} and \textit{njs} of the package \textit{ape} in the R environment. These two techniques are explained in Appendix~\ref{AppendixMethodsNJ} and \ref{AppendixMethodsKM}. The reader should refer to the following papers for more details on these methods: \citet{kmeans1967,kmeans2010,NJ1987,NJ2006,Blanco-Cuaresma2015,Fraix-Burnet2015,Jofre2017}.

\subsection{Selection of the chemical elements}
\label{sec:param_selection}

As will be shown, the result obtained by the MP technique with all the derived abundances for all the analyzed chemical elements is already quite satisfactory. We however tried to improve it by selecting a subset of the chemical elements since, as explained in the Appendix~\ref{AppendixMethods}, the success of phylogenetic analyses rely on the descriptors being good tracers of the history of the Galaxy. For this purpose we used several tools, like Principal Component Analysis, Independent Component Analysis, Multivariate Clustering of input variables, and many trial and error runs. Note that these tools allow a better understanding of the data, but they do not tell anything on the phylogenetic interest of the abundances. Only trials and errors MP analyses give the correct information, or, ideally, some theoretical arguments that we presently lack.

At the end, our best subset is made of eight elements: Al Ba Co Fe Mg Mn Sc Ti. It is the best we have found, we cannot exclude that there could be still a better one. Note that we have found another subset which yields very close results: Ba C Ca Fe Mg Mn Si Ti Y. 

In this paper, we thus consider three samples (Table~\ref{TabAbundances}): Sfull (calibrated from the Sun with all 25 abundances), Mfull and Msel (calibrated from M67 with respectively 29 and 8 abundances).

\section{Results}
\label{results}

\subsection{MP Analysis}
\label{cladresult}

   \begin{figure*}
   \centering
   \includegraphics[width=0.32\linewidth]{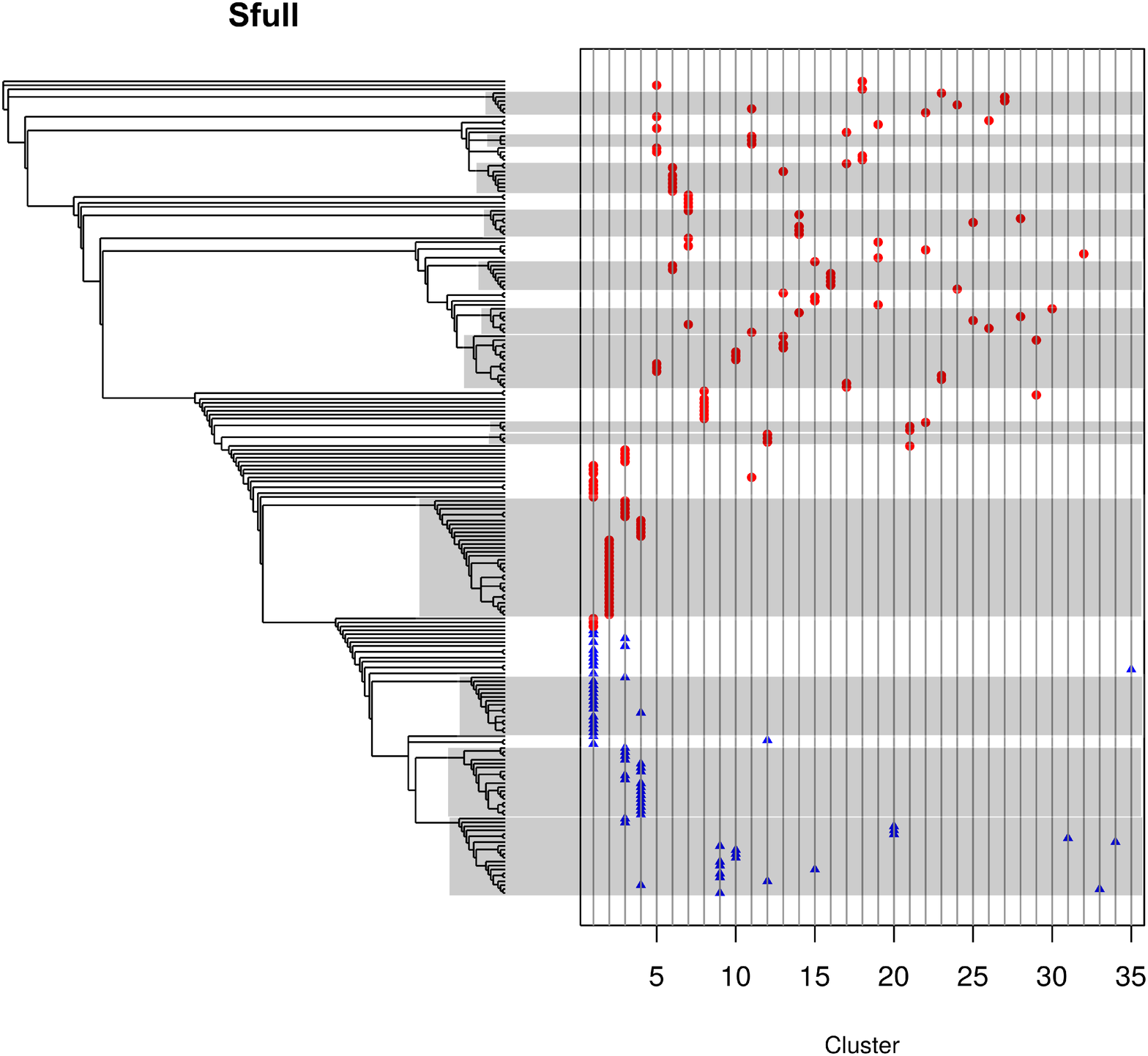} 
    \includegraphics[width=0.32\linewidth]{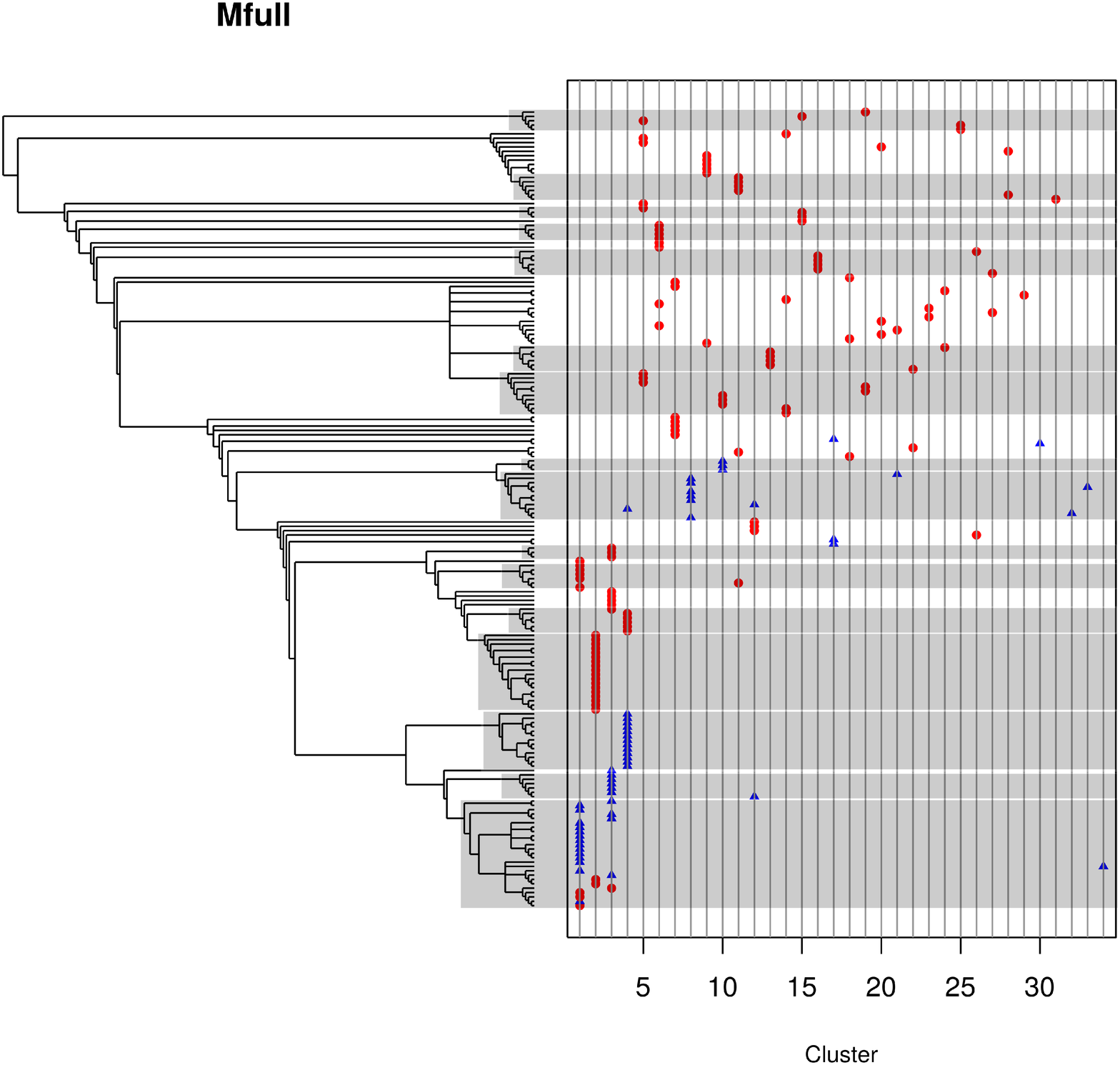} 
     \includegraphics[width=0.30\linewidth]{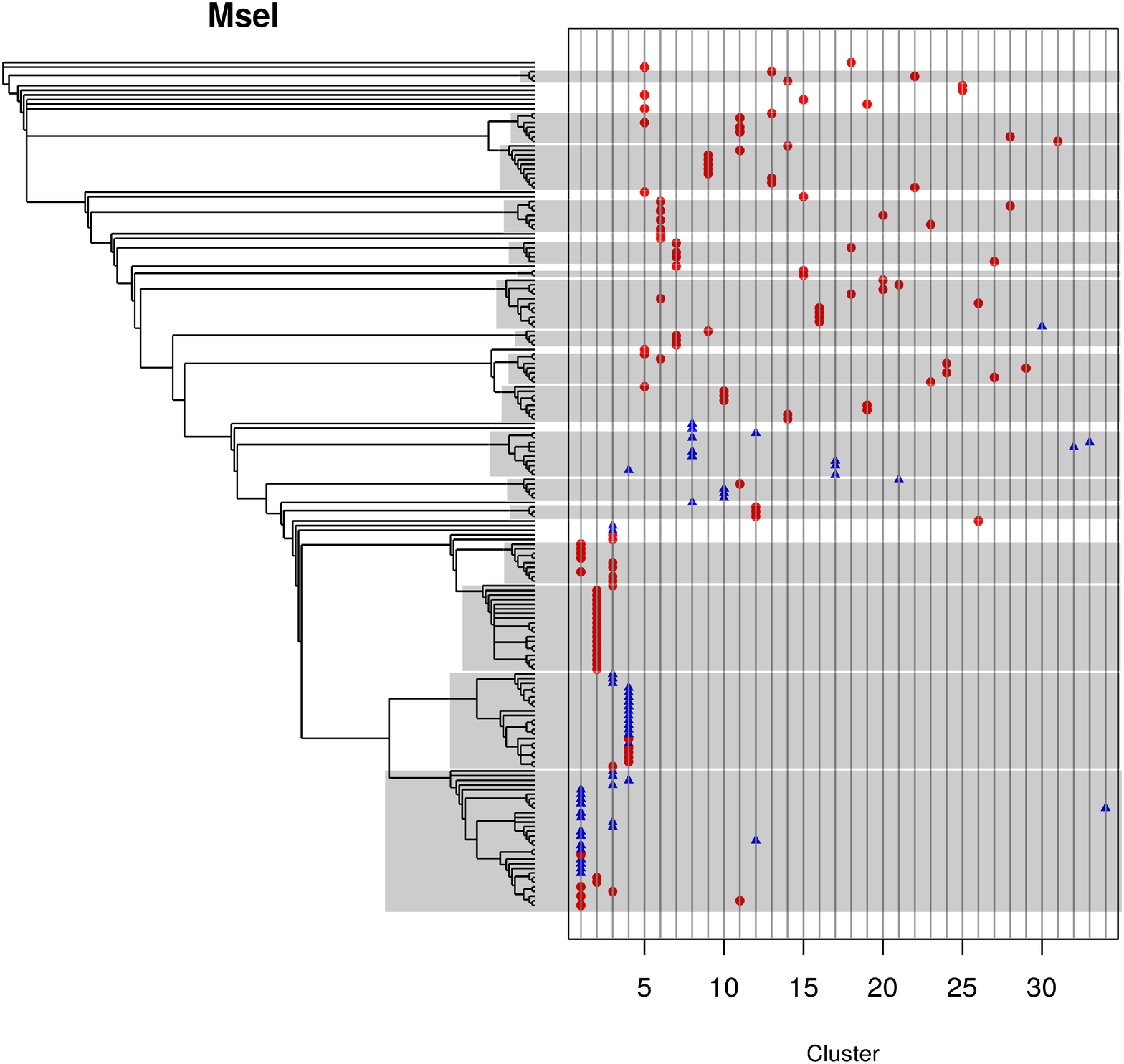}
   \caption{Cladograms obtained for 207 stars from 34 clusters calibrated using the Sun and with abundances from 25 elements (left), for 180 stars from 33 clusters calibrated using M67 with abundances from 29 elements (middle) and with abundances from 8 elements (right). Red points and blue triangles correspond respectively to red giant and dwarf stars. The horizontal axis gives the stellar cluster index as given in  Table~\ref{TabClusters}, in decreasing order of the number of stars. The gray boxes indicate the structures on the tree that could be defined as groups. They also here serve as a visual aid: the more the stars are stacked together vertically and the better they belong to the same gray box, the better is the phylogenetic reconstruction of the stellar clusters.}
              \label{FigTrees}%
    \end{figure*}

The results of the MP analyses of Sfull, Mfull and Msel are shown on Fig.~\ref{FigTrees}. Dwarfs and red giants are shown with given different colors and symbols, and gray boxes underline sub-structures of the tree that are identified to the families of stars. The open clusters are ranked on the abscissa in decreasing number of stars as given in Table~\ref{TabClusters}. Ideally, for a given open cluster, all the stars should be stacked together vertically, and correspond to a unique sub-structure on the tree.

On the Sfull tree (Fig.~\ref{FigTrees}, left), the dwarf and the red giant stars are perfectly separated irrespective of the true families. Due to the gap in our sample between the dwarf and red giants stars, even in the abundance space, the MP analysis naturally finds that the shortest route is to connect first each category separately. Even if some families are partially gathered in few groups, it would be difficult in practice to recognize that these groups indeed belong to the same family. This result is thus not satisfactory. 

Performing differential abundance analysis for dwarfs and giants separately improves the result very significantly (Mfull, Fig.~\ref{FigTrees}, middle) because families much better match sub-structures on the tree (the differences due to different evolutionary stages mentioned before are minimized). For all clusters, dwarfs and giants are better mixed up, even if this is not perfect (see for instance cluster 3 which is \object{IC4651}). The differential calibration considering the evolutionary stage of the stars thus seem a necessary procedure.

The selection of elements (Msel, Fig.~\ref{FigTrees}, right) still improves the result: this is almost perfect for the clusters 2 and 4, and for the largest part of cluster 1. Cluster 3 is more problematic and we never managed to get a better result for it. Interestingly, as can be already noticed on the Mfull result and much more clearly on the Msel one, this cluster (IC4651) is mixed with cluster 1 (M67) on the same sub-structures of the tree. This means that we are not able to separate this two open clusters, indicating that they probably have very similar chemical compositions (see Sect.~\ref{discussion}). Many smaller groups are well reconstructed as well, clearly better than for Mfull.

To quantify the quality of the classifications and compare them, we present in Appendix~\ref{TabContingencies} contingency tables with recall and precision estimators for the four biggest open clusters M67, NGC6705, IC4651 and NGC2632. The other clusters have less than 8 members so that these indicators are not very much reliable. The recall (or sensitivity) is the fraction of stars of an open cluster (true class) that are retrieved in the same group. The precision is the fraction of stars of a group that are members of a same open cluster. Since we are interested in reconstituting families of stars belonging to the same open cluster, our main criterion is the recall estimator.

There are more stars of the biggest open cluster outside the defined groups (situated at individual branches) for Sfull (Table~\ref{TabContSfull}) than for Msel (Table~\ref{TabContMsel}) or Mfull (Table~\ref{TabContMfull}) because in this case (Sfull) the chemical abundances were not computed to minimize differences due to the evolutionary stages (dwarfs and giants from the same open clusters end up in different families as shown in the Sfull tree in Fig.~\ref{FigTrees}). This is reflected in the recall values which are higher for Msel. The precision is more difficult to interpret. For instance the group corresponding to the open cluster M67 contains more stars of other open clusters (not shown in the Tables) in Msel than in Sfull, despite it has most of its stars in Msel. However, the precision looks slightly better for Msel than Sfull, and Mfull seems even slightly better. Globally, we want to maximize recall and precision values and Msel seems to have the best balance.

\subsection{NJ result}
\label{njresult}

\begin{figure}[t]
\centering
\includegraphics[width=0.8\linewidth]{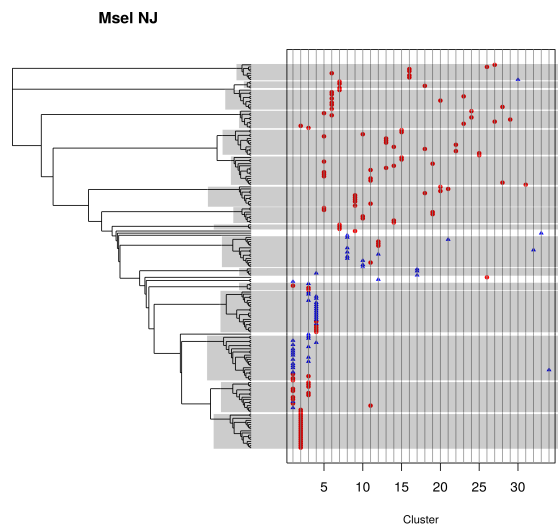} 
\caption{Tree obtained with the Neighbor Joining Tree Estimation method for 183 stars from 33 clusters calibrated using M67 with abundances from 8 elements. Otherwise same as Fig.~\ref{FigTrees}.}
\label{FigTreeNJ}%
\end{figure}

We have performed a Neighbor-Joining analysis to test the robustness of the presence of a phylogenetic signal. Theoretically, when the phylogeny is perfect, the trees from NJ and MP are identical \citep[e.g. ][]{Fraix-Burnet2015}. Even though this ideal situation is never met in real life, at least this is a safe check. We show in Fig.~\ref{FigTreeNJ} only the Neighbor-Joining result for Msel since the selected elements are supposedly the most suited to reconstruct the phylogeny.  

The NJ family reconstruction is rather good, quite comparable to the MP tree for Msel (Fig.~\ref{FigTrees}). Maybe the biggest cluster M67 is better reconstructed in the MP analysis, otherwise this is quite globally similar for the smaller groups. This is confirmed by the contingency table (Table~\ref{TabContMselNJ}) that is similar to the one for Msel (Table~\ref{TabContMsel}) with very close recall and precision values with a significantly higher recall for M67 in Msel.  This indicates that the trees are very close to each other, giving strong support to the MP analysis and more importantly, confirms that the selected elements bear a real phylogenetic signal.

\subsection{K-means result}
\label{kmresult}

\begin{figure}[t]
\centering
\includegraphics[width=0.8\linewidth]{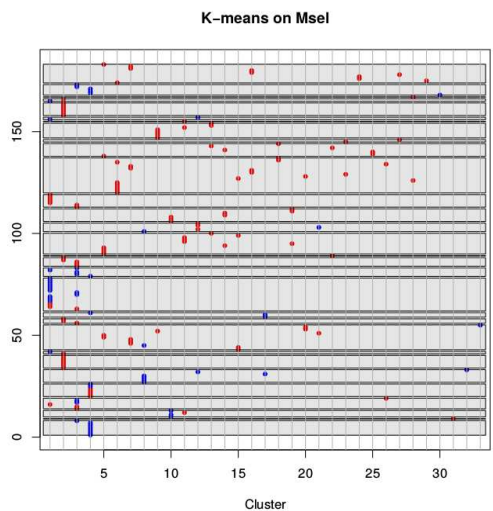} 
\caption{K-means correspondence plot for 30 groups (Sect~\ref{kmresult}) in the same presentation as the other figures. The vertical order of the kmeans groups represented by the gray boxes is arbitrary.}
\label{Fig:kmeans30}
\end{figure}

We have also also performed a k-means analysis for Msel. The big problem of this partitioning technique is the a priori choice of the number of clusters, which in practice is unknown. Another problem is that it cannot take unknown values, hence we replaced two unknown Al abundances in the Msel dataset by the mean of this parameter in this sample. We have used the package \textit{NbClust} from the R environment in which many techniques can be used to infer the "best" number of groups present in the data. The majority of the tests find either 2 or 30 as the optimal number of groups. We have chosen 30 groups, that is remarkably close to the real number of open clusters, but somewhat surprising due to the very small number of stars in many of them. We show the correspondence between the groups and the true families in Fig.~\ref{Fig:kmeans30} in a similar way as the other figures except that the vertical order of the k-means clusters is arbitrary.  

None of the biggest open clusters are correctly reconstructed, i.e. each family is split in several k-means groups, confirming that partitioning techniques are not adapted to the chemical tagging objectives, even with the subset of elements selected on a phylogenetic basis. The contingency table (Table~\ref{TabContMselkm}) has more groups (15) than all other results except for Sfull, and the recall values are clearly the worst of all. The precision values seem high, but NGC6705 and NGC2632 cannot be identified to a single group (groups 2 and 3 for the first cluster, groups 5 and 6 for the second one).

\section{Discussion}
\label{discussion}

The two different phylogenetic techniques agree with each other and yield a very satisfactory reconstruction, especially for the biggest stellar clusters. In both cases, there is some arbitrariness in the precise definition of the families (particularly for the smallest clusters, where no robust quantitative estimates are possible). However, we remained rather conservative and consider only visually obvious sub-structures of the trees. In a real scenario, a subsequent analysis, not done in this paper, of the kinematics of the stars would provide significant information to assess and justify the results.

We conclude that the phylogenetic approaches can essentially improve the chemical tagging family reconstruction provided calibration of dwarfs and red giants are made properly and a proper element selection is done. We discuss here on possible explanations for the chemical element subset we have found.

We have shown that the selection of the chemical elements, being in any case a necessary step for phylogenetic analyses, improves considerably the reconstruction of the stellar families. We have found by trials and errors a very satisfactory subset of eight elements. An important question is: why this subset?

The first idea is that this is an artifact from the data. In particular, the abundances for these elements could be more reliable because they have lower uncertainties. This is however not true for our sample. Inspecting the standard deviation per cluster for this selection of elements (Table~\ref{tab:abundances_summary_msel}) shows a low dispersion that can also be found in some of the non-selected elements (Table~\ref{tab:abundances_summary_rest1} and Table~\ref{tab:abundances_summary_rest2}).

The second possibility is that these elements are less affected by stellar evolution, in the sense that they better probe the true chemical composition of the star at the time of observation. But when we verify the relation between the abundances and the evolutionary stage using the Spearman rank-order correlation coefficient, the results for this set of elements are not particularly different from the rest of the abundances.

The third and more exciting possibility is that, based on the differential approach with respect to M67 that we followed, these elements are better tracers of subsequent generation of stars, they could constitute the equivalent of the DNA of living organisms used to build the 'Tree of Life' \citep[e.g. ][]{Dress2010,Vienne2016}. We do not claim that this is the absolute best combination of elements, future works should apply this phylogenetic analysis to bigger datasets of stars and chemical elements (improving data quality and analysis techniques when possible). There is a delicate equilibrium between how much phylogenetic information a chemical element provides and how much noise it introduces due to difficulties measuring its abundance from stellar spectra. Some of the selected elements could be replaced by counterparts with which they are correlated and for which more precise abundance measurements are possible, some other elements could be included if the abundance determination is improved. These factors have the potential to reshape the selection of elements and, for instance, in Sect.~\ref{sec:param_selection} we already indicate that a different set of elements yields close results.

This subset of elements cover difference nucleosynthetic processes. Iron (mainly created via explosive nucleosynthesis) is a key element that traditionally has been used to represent the overall metallicity evolution of stars. It is expected to be part of the selected elements since it contains a lot of information and the rest of abundances have been expressed relative to it. All the elements except Aluminium and Barium (i.e., Fe, Sc, Mn, Co, Ti, Mg) are fundamentally synthesized during the core-collapse phase of a core-collapse massive star supernova or during a thermonuclear explosion involving a less massive star. Aluminium is mainly synthesized in the interior of the stars from Ne and C burning. Barium is the only element in this group that is synthesized due to neutron capture processes such as s-processes (mainly) and r-processes \citep{1989GeCoA..53..197A, 1999ApJ...525..886A}. It is expected that good proxies for the history of the Galaxy should contain a set of elements with decoupled synthesis rates (as it is the case for this subset) because it maximizes the information needed to distinguish between families of stars.

When comparing to the second subset with close results (Sect.~\ref{sec:param_selection}), both groups have Ba Fe Mg Mn Si Ti in common and Co Al are replaced by C Ca Y. Cobalt is one of the elements mainly created by explosive nucleosynthesis as Calcium also is. Aluminium is synthesized in the interior of the stars but Carbon also is. Finally, Yttrium is a neutron capture element and that nucleosynthetic process is already represented by Barium. This confirms the idea that other subsets of elements may be found with slight variations and this works demonstrates how phylogenetic techniques can be used to identify the best set.

The quality of the data and the uncertainties naturally limit our ability to distinguish families. In our sample, the uncertainties are sometimes as large as the spread within the sample. Our result is remarkably good taking this into consideration, and this might indicate that phylogenetic approaches may be more tolerant to uncertainties.

Another important limitation to chemical tagging can appear if stars born from the same molecular cloud do not share a unique abundance pattern as suggested by \cite{Smiljanic2017}, based on an experiment where a hierarchical clustering technique was used. If we assume that this hypothesis is true, then it seems that our phylogenetic approach is less sensitive to this level of abundance discrepancies, or the open clusters in our sample happen to be sufficiently homogeneous and the spectroscopic analysis succeeded in reducing evolutionary effects by using a differential approach with respect to M67.

Several distinct molecular clouds may also have similar initial chemical composition. As a consequence, from the phylogenetic point of view, a family of stars does not necessarily mean a single open cluster. For instance, this could be the reason why we were not able to reconstruct correctly the group IC~4651, that appears to be mixed with M67 on our Msel tree (Fig.~\ref{FigTrees}). It happens that the chemical signature of IC~4651 was already identified as very similar to the one of M67 \citep{2016sf2a.conf..333B}. This is probably not an issue for the study of the chemical history of star formation within our Galaxy, but this is definitively one if we are also interested by the spatial origin of these stars. In the latter case, it would be necessary to include dynamical studies based on spatial positions and motions in conjunction to the phylogenetic analysis based on abundances.

Some of the previous limitations could explain the difficulties for some of the smaller clusters, but their very low statistics prevents any further investigation. This raises again the importance of future phylogenetic analysis with a greater number of stars.

\section{Conclusion}
\label{conclusion}

We have shown that phylogenetic analysis is a very promising tool to reconstitute families of stars born from similar molecular clouds (i.e., chemical tagging). Based on the requirement by this technique that the input variables (i.e., chemical abundances) must be good tracers of the history of the Galaxy, we conclude that a careful selection of the chemical elements must be made. We show that this improves considerably the quality of the result, and we also proved that using a differential spectroscopic analysis with respect to an open cluster (in particular for this study, a dwarf and a giant star from M67) is a better strategy than the traditional approach of using the Sun as reference.

Questions arise on why the subset of chemical elements we have found is well suited for the phylogenetic chemical tagging study. We explored the idea that they may be intrinsic marks of families of stars, being somewhat a proxy for the "DNA of stars". Other good proxies could be eventually found with some variations on the selected elements, further investigation will be needed using a greater sample of stars. 

This works opens a new approach to study the history of star formation in our Galaxy based on the stellar chemical signatures.

\begin{acknowledgements}
This work would not have been possible without the support of Dr. Laurent Eyer (University of Geneva). The authors thank Paula Jofr\'e for contributing with interesting different points of view regarding chemical tagging and phylogenetic studies, and Carine Babusiaux for interesting discussions. We thank the anonymous referee for suggesting significant improvements to a first version of this paper. This research has made use of NASA's Astrophysics Data System.
 \end{acknowledgements}

% WARNING
%-------------------------------------------------------------------
% Please note that we have included the references to the file aa.dem in
% order to compile it, but we ask you to:
%
% - use BibTeX with the regular commands:
   \bibliographystyle{aa} % style aa.bst
   \bibliography{phylochemtag} % your references Yourfile.bib
%
% - join the .bib files when you upload your source files
%-------------------------------------------------------------------

%\listofobjects

\onllongtab{
%\begin{landscape}
\footnotesize  % Switch from 12pt to 11pt; otherwise, table won't fit
\begin{longtable}{l|cc|cc|cc|cc|cc|cc|cc|cc}
\caption{Differential abundances with respect to M67 for the 'Msel' set of elements} \label{tab:abundances_summary_msel}\\
 & \multicolumn{2}{c|}{[Fe/H]} & \multicolumn{2}{c|}{[Sc/Fe]} & \multicolumn{2}{c|}{[Mn/Fe]} & \multicolumn{2}{c|}{[Co/Fe]} & \multicolumn{2}{c|}{[Ti/Fe]} & \multicolumn{2}{c|}{[Mg/Fe]} & \multicolumn{2}{c|}{[Al/Fe]} & \multicolumn{2}{c}{[Ba/Fe]} \\
 & $\bar{x}$    &   $\sigma$  & $\bar{x}$    &   $\sigma$  & $\bar{x}$    &   $\sigma$  & $\bar{x}$    &   $\sigma$  & $\bar{x}$    &   $\sigma$  & $\bar{x}$    &   $\sigma$  & $\bar{x}$    &   $\sigma$  & $\bar{x}$    &   $\sigma$  \\
\hline
\footnotesize{Collinder350} & 0.03 &  & 0.00 &  & -0.03 &  & -0.11 &  & -0.02 &  & -0.07 &  & -0.08 &  & 0.47 &  \\
\footnotesize{IC2714} & -0.02 & 0.04 & -0.02 & 0.01 & -0.04 & 0.02 & -0.10 & 0.01 & -0.04 & 0.02 & -0.06 & 0.01 & -0.06 & 0.01 & 0.34 & 0.05 \\
\footnotesize{IC4651} & 0.04 & 0.04 & -0.01 & 0.02 & -0.01 & 0.02 & -0.03 & 0.02 & -0.02 & 0.01 & -0.04 & 0.03 & -0.03 & 0.04 & 0.03 & 0.04 \\
\footnotesize{IC4756} & -0.01 & 0.01 & -0.03 & 0.01 & -0.07 & 0.01 & -0.08 & 0.04 & -0.02 & 0.03 & -0.04 & 0.01 & -0.07 & 0.02 & 0.26 & 0.04 \\
\footnotesize{M67} & -0.02 & 0.03 & 0.00 & 0.02 & -0.01 & 0.02 & -0.01 & 0.01 & -0.00 & 0.02 & -0.01 & 0.03 & -0.01 & 0.04 & 0.00 & 0.05 \\
\footnotesize{Melotte111} & -0.05 & 0.03 & -0.04 & 0.02 & -0.03 & 0.01 & -0.07 & 0.01 & 0.00 & 0.01 & -0.04 & 0.02 & -0.05 & 0.04 & 0.26 & 0.05 \\
\footnotesize{Melotte20} & 0.06 &  & -0.04 &  & -0.09 &  & -0.05 &  & -0.07 &  & -0.01 &  & -0.15 &  & 0.22 &  \\
\footnotesize{Melotte22} & -0.01 &  & -0.08 & 0.02 & 0.00 & 0.04 & -0.06 & 0.02 & -0.02 & 0.03 & -0.01 & 0.01 & -0.02 &  & 0.20 & 0.04 \\
\footnotesize{Melotte71} & -0.17 & 0.07 & -0.03 & 0.04 & -0.12 & 0.04 & -0.16 & 0.11 & -0.03 & 0.01 & -0.05 & 0.04 & -0.02 & 0.03 & 0.47 & 0.04 \\
\footnotesize{NGC1817} & -0.15 & 0.02 & -0.05 & 0.02 & -0.06 & 0.02 & -0.11 & 0.01 & -0.04 & 0.02 & -0.10 & 0.07 & -0.04 & 0.03 & 0.40 & 0.03 \\
\footnotesize{NGC2099} & 0.07 & 0.02 & -0.02 & 0.01 & 0.08 & 0.09 & -0.12 & 0.01 & -0.04 & 0.01 & -0.14 & 0.08 & -0.10 & 0.04 & 0.39 & 0.11 \\
\footnotesize{NGC2251} & 0.05 & 0.03 & -0.01 & 0.02 & -0.03 & 0.01 & -0.11 &  & -0.03 & 0.01 & -0.08 & 0.06 & -0.07 & 0.01 & 0.43 & 0.04 \\
\footnotesize{NGC2287} & -0.10 & 0.01 & -0.01 & 0.01 & -0.05 &  & -0.08 & 0.01 & -0.06 & 0.01 & -0.04 & 0.01 & -0.04 &  & 0.34 & 0.01 \\
\footnotesize{NGC2360} & -0.06 & 0.02 & -0.05 & 0.01 & -0.07 & 0.02 & -0.12 & 0.01 & -0.04 & 0.01 & -0.12 & 0.02 & -0.07 & 0.01 & 0.30 & 0.01 \\
\footnotesize{NGC2423} & 0.04 & 0.05 & -0.02 & 0.01 & -0.03 &  & -0.07 & 0.02 & -0.04 & 0.01 & -0.06 & 0.03 & -0.06 & 0.04 & 0.15 & 0.03 \\
\footnotesize{NGC2447} & -0.09 & 0.06 & -0.02 & 0.01 & -0.08 & 0.04 & -0.07 & 0.03 & -0.00 & 0.02 & -0.05 & 0.03 & -0.05 & 0.03 & 0.34 & 0.06 \\
\footnotesize{NGC2477} & 0.07 &  & -0.05 &  & -0.05 &  & -0.13 &  & -0.09 &  & -0.16 &  & -0.19 &  & 0.26 &  \\
\footnotesize{NGC2539} & 0.01 & 0.02 & -0.02 & 0.01 & -0.01 & 0.02 & -0.10 & 0.00 & -0.04 & 0.01 & -0.11 & 0.03 & -0.08 & 0.01 & 0.28 & 0.05 \\
\footnotesize{NGC2547} & -0.08 &  & -0.06 &  & 0.01 &  & -0.07 &  & -0.01 &  & -0.05 &  & -0.04 &  & 0.20 &  \\
\footnotesize{NGC2548} & 0.05 & 0.04 & -0.01 & 0.01 & 0.02 & 0.09 & -0.09 & 0.01 & -0.04 & 0.02 & -0.08 & 0.01 & -0.07 &  & 0.39 & 0.15 \\
\footnotesize{NGC2567} & -0.05 & 0.06 & -0.03 & 0.01 & -0.08 & 0.07 & -0.10 & 0.02 & -0.03 & 0.02 & -0.10 & 0.02 & -0.06 & 0.02 & 0.38 & 0.06 \\
\footnotesize{NGC2632} & 0.15 & 0.04 & -0.02 & 0.02 & 0.02 & 0.02 & -0.04 & 0.03 & -0.03 & 0.01 & -0.05 & 0.02 & -0.08 & 0.04 & 0.08 & 0.04 \\
\footnotesize{NGC3114} & -0.02 & 0.06 & -0.03 & 0.02 & -0.06 & 0.02 & -0.09 & 0.03 & -0.04 & 0.01 & -0.08 & 0.05 & -0.05 & 0.04 & 0.24 & 0.07 \\
\footnotesize{NGC3532} & 0.00 & 0.06 & -0.03 &  & -0.09 & 0.04 & -0.09 & 0.01 & -0.04 & 0.02 & -0.08 & 0.02 & -0.06 & 0.02 & 0.37 & 0.04 \\
\footnotesize{NGC3680} & -0.04 & 0.04 & -0.01 & 0.02 & -0.06 & 0.02 & -0.05 & 0.01 & 0.01 & 0.01 & -0.04 & 0.05 & -0.04 & 0.04 & 0.12 & 0.03 \\
\footnotesize{NGC4349} & -0.11 & 0.04 & -0.02 & 0.04 & -0.05 & 0.01 & -0.09 & 0.05 & -0.05 & 0.04 & -0.07 & 0.05 & -0.04 & 0.08 & 0.31 & 0.17 \\
\footnotesize{NGC5822} & -0.04 & 0.01 & -0.07 & 0.05 & -0.04 & 0.01 & -0.12 & 0.02 & -0.07 & 0.01 & -0.19 & 0.07 & -0.10 & 0.03 & 0.58 & 0.44 \\
\footnotesize{NGC6475} & 0.03 &  & -0.05 &  & -0.02 &  & -0.07 &  & -0.01 &  & -0.09 &  & -0.07 &  & 0.17 &  \\
\footnotesize{NGC6494} & -0.03 & 0.02 & -0.03 & 0.01 & -0.02 & 0.01 & -0.11 & 0.02 & -0.04 & 0.02 & -0.07 & 0.02 & -0.05 & 0.01 & 0.33 & 0.03 \\
\footnotesize{NGC6633} & 0.01 & 0.01 & -0.04 & 0.02 & -0.04 & 0.01 & -0.11 & 0.01 & -0.04 & 0.01 & -0.09 & 0.04 & -0.09 & 0.01 & 0.30 & 0.03 \\
\footnotesize{NGC6705} & 0.04 & 0.04 & 0.02 & 0.02 & 0.06 & 0.02 & -0.03 & 0.02 & -0.02 & 0.02 & -0.01 & 0.05 & -0.00 & 0.03 & 0.14 & 0.05 \\
\footnotesize{NGC6940} & 0.11 & 0.03 & -0.03 & 0.01 & -0.03 & 0.01 & -0.10 & 0.01 & -0.05 & 0.02 & -0.09 & 0.02 & -0.08 & 0.01 & 0.26 & 0.02 \\
\footnotesize{NGC752} & 0.01 & 0.01 & -0.02 & 0.01 & -0.05 & 0.02 & -0.10 & 0.01 & -0.02 & 0.01 & -0.10 & 0.02 & -0.05 & 0.01 & 0.26 & 0.02 \\
\end{longtable}
\tablefoot{Empty values represent lack of enough suitable absorption lines and/or stars to compute the statistic.}
%\end{landscape}
}% End onllongtab

\onllongtab{
\begin{landscape}
\footnotesize  % Switch from 12pt to 11pt; otherwise, table won't fit
\begin{longtable}{l|cc|cc|cc|cc|cc|cc|cc|cc|cc|cc|cc}
\caption{Differential abundances with respect to M67 for the rest of elements not included in 'Msel'} \label{tab:abundances_summary_rest1}\\
 & \multicolumn{2}{c|}{[C/Fe]} & \multicolumn{2}{c|}{[Ca/Fe]} & \multicolumn{2}{c|}{[Ce/Fe]} & \multicolumn{2}{c|}{[Cr/Fe]} & \multicolumn{2}{c|}{[Cu/Fe]} & \multicolumn{2}{c|}{[Eu/Fe]} & \multicolumn{2}{c|}{[La/Fe]} & \multicolumn{2}{c|}{[Mo/Fe]} & \multicolumn{2}{c|}{[Na/Fe]} & \multicolumn{2}{c|}{[Nd/Fe]} & \multicolumn{2}{c}{[Ni/Fe]} \\
 & $\bar{x}$    &   $\sigma$  & $\bar{x}$    &   $\sigma$  & $\bar{x}$    &   $\sigma$  & $\bar{x}$    &   $\sigma$  & $\bar{x}$    &   $\sigma$  & $\bar{x}$    &   $\sigma$  & $\bar{x}$    &   $\sigma$  & $\bar{x}$    &   $\sigma$  & $\bar{x}$    &   $\sigma$  & $\bar{x}$    &   $\sigma$  & $\bar{x}$    &   $\sigma$  \\
\hline
\footnotesize{Collinder350} & -0.09 &  & 0.04 &  & 0.15 &  & -0.02 &  & -0.20 &  & 0.02 &  & 0.21 &  &  &  & -0.06 &  & 0.16 &  & -0.10 &  \\
\footnotesize{IC2714} & -0.10 & 0.06 & 0.04 & 0.01 & 0.11 & 0.04 & -0.02 & 0.01 & -0.14 & 0.04 & -0.02 & 0.02 & 0.23 & 0.04 & 0.11 & 0.04 & 0.02 & 0.07 & 0.10 & 0.03 & -0.08 & 0.01 \\
\footnotesize{IC4651} & 0.01 & 0.04 & 0.03 & 0.03 & 0.02 & 0.08 & 0.00 & 0.02 & -0.05 & 0.05 & -0.09 & 0.03 & -0.02 & 0.08 & -0.04 & 0.04 & 0.00 & 0.05 & -0.02 & 0.03 & -0.01 & 0.02 \\
\footnotesize{IC4756} & -0.07 & 0.06 & 0.06 & 0.02 & 0.15 & 0.02 & -0.01 & 0.04 & -0.11 & 0.01 & -0.03 &  & 0.12 & 0.08 & 0.10 &  & -0.07 & 0.05 & 0.15 & 0.01 & -0.07 & 0.03 \\
\footnotesize{M67} & 0.04 & 0.07 & 0.03 & 0.03 & -0.00 & 0.06 & -0.01 & 0.01 & -0.02 & 0.04 & -0.05 & 0.02 & -0.05 & 0.06 & -0.04 & 0.03 & 0.02 & 0.03 & -0.01 & 0.03 & -0.01 & 0.01 \\
\footnotesize{Melotte111} & -0.04 & 0.08 & 0.06 & 0.04 & 0.14 & 0.07 & 0.03 & 0.01 & -0.12 & 0.02 &  &  & 0.15 & 0.05 &  &  & -0.05 & 0.05 & 0.12 & 0.06 & -0.06 & 0.01 \\
\footnotesize{Melotte20} & -0.10 &  & 0.07 &  & 0.10 &  & 0.00 &  & -0.23 &  &  &  &  &  &  &  & -0.12 &  & 0.08 &  & -0.06 &  \\
\footnotesize{Melotte22} & -0.03 & 0.03 & 0.02 & 0.02 &  &  & 0.00 & 0.01 & -0.07 & 0.08 &  &  &  &  &  &  & -0.04 & 0.04 & 0.02 &  & 0.00 & 0.02 \\
\footnotesize{Melotte71} & -0.06 & 0.06 & 0.03 & 0.01 & 0.12 & 0.04 & -0.08 & 0.06 & -0.31 & 0.20 & 0.04 & 0.01 & 0.33 & 0.06 & 0.08 &  & -0.04 & 0.19 & 0.14 & 0.01 & -0.11 & 0.04 \\
\footnotesize{NGC1817} & -0.14 & 0.10 & 0.06 & 0.02 & 0.17 & 0.03 & -0.07 & 0.01 & -0.19 & 0.08 & 0.04 & 0.07 & 0.32 & 0.04 & 0.20 & 0.07 & -0.09 & 0.03 & 0.17 & 0.02 & -0.08 & 0.02 \\
\footnotesize{NGC2099} & 0.12 & 0.21 & -0.03 &  & -0.00 & 0.12 & 0.07 & 0.12 & -0.16 & 0.11 & 0.10 & 0.13 & 0.17 & 0.06 & 0.30 & 0.12 & -0.04 &  & 0.21 & 0.15 & -0.08 & 0.01 \\
\footnotesize{NGC2251} & -0.18 & 0.09 & 0.01 & 0.01 & 0.16 & 0.02 & -0.02 & 0.01 & -0.17 & 0.04 & 0.03 & 0.02 & 0.33 &  & 0.14 & 0.06 & -0.04 & 0.02 & 0.13 & 0.01 & -0.08 & 0.01 \\
\footnotesize{NGC2287} & -0.03 & 0.04 & 0.01 & 0.01 & 0.10 & 0.01 & -0.04 & 0.01 & -0.16 &  & 0.01 & 0.01 & 0.35 & 0.01 & 0.03 & 0.01 & 0.06 & 0.01 & 0.09 & 0.01 & -0.08 &  \\
\footnotesize{NGC2360} & -0.07 & 0.05 & 0.07 & 0.01 & 0.10 & 0.04 & -0.05 & 0.01 & -0.13 & 0.02 & -0.02 & 0.02 & 0.19 & 0.02 & 0.11 & 0.02 & -0.04 & 0.02 & 0.10 & 0.03 & -0.09 & 0.00 \\
\footnotesize{NGC2423} & -0.09 & 0.12 & 0.01 & 0.02 & 0.04 & 0.02 & -0.01 & 0.02 & -0.07 & 0.02 & -0.04 & 0.01 & 0.11 & 0.09 & 0.04 & 0.05 & 0.01 & 0.01 & 0.03 & 0.01 & -0.06 & 0.01 \\
\footnotesize{NGC2447} & -0.15 & 0.15 & 0.06 & 0.01 & 0.24 & 0.03 & -0.01 & 0.02 & -0.11 & 0.04 & 0.01 & 0.03 & 0.18 & 0.07 & 0.09 & 0.03 & -0.05 & 0.04 & 0.19 & 0.05 & -0.08 & 0.02 \\
\footnotesize{NGC2477} & -0.18 &  & 0.05 &  & -0.09 &  & -0.04 &  & -0.29 &  & -0.09 &  & 0.02 &  &  &  & 0.04 &  & 0.03 &  & -0.07 &  \\
\footnotesize{NGC2539} & -0.06 & 0.06 & 0.03 & 0.01 & 0.06 & 0.03 & -0.02 & 0.01 & -0.12 & 0.03 & -0.06 & 0.02 & 0.15 & 0.03 & 0.13 & 0.05 & 0.02 & 0.04 & 0.08 & 0.03 & -0.07 & 0.01 \\
\footnotesize{NGC2547} &  &  & 0.11 &  &  &  & 0.03 &  & -0.10 &  &  &  &  &  &  &  & 0.01 &  & 0.15 &  & -0.03 &  \\
\footnotesize{NGC2548} & 0.04 & 0.11 & 0.01 & 0.01 & 0.10 & 0.03 & 0.01 & 0.04 & -0.12 & 0.04 & 0.03 &  & 0.17 & 0.01 & 0.10 &  & -0.01 & 0.09 & 0.15 & 0.04 & -0.07 & 0.02 \\
\footnotesize{NGC2567} & -0.21 & 0.20 & 0.05 & 0.02 & 0.18 & 0.09 & -0.02 & 0.02 & -0.14 & 0.07 & -0.00 & 0.05 & 0.25 & 0.04 & 0.12 & 0.06 & -0.00 & 0.04 & 0.15 & 0.05 & -0.10 & 0.01 \\
\footnotesize{NGC2632} & -0.04 & 0.05 & 0.02 & 0.02 & -0.02 & 0.04 & -0.01 & 0.02 & -0.05 & 0.07 & -0.08 & 0.01 & -0.00 & 0.06 & 0.06 & 0.01 & 0.05 & 0.07 & 0.01 & 0.12 & -0.01 & 0.02 \\
\footnotesize{NGC3114} & -0.10 & 0.10 & 0.03 & 0.04 & 0.11 & 0.04 & -0.03 & 0.03 & -0.12 & 0.03 & -0.01 & 0.01 & 0.22 & 0.10 & 0.06 & 0.07 & 0.06 & 0.10 & 0.08 & 0.04 & -0.08 & 0.02 \\
\footnotesize{NGC3532} & -0.20 & 0.18 & 0.03 & 0.02 & 0.21 & 0.13 & -0.01 & 0.01 & -0.21 & 0.08 & 0.00 & 0.01 & 0.25 & 0.07 & 0.10 & 0.01 & 0.01 & 0.07 & 0.10 & 0.01 & -0.08 &  \\
\footnotesize{NGC3680} & -0.01 & 0.08 & 0.04 & 0.02 & 0.15 & 0.06 & -0.00 & 0.01 & -0.08 & 0.03 & 0.02 & 0.02 & 0.12 & 0.14 & 0.00 & 0.02 & -0.07 & 0.03 & 0.08 & 0.06 & -0.05 & 0.01 \\
\footnotesize{NGC4349} & -0.10 & 0.15 & 0.05 & 0.02 & 0.07 & 0.06 & -0.03 & 0.03 & -0.12 & 0.12 & 0.02 & 0.03 & 0.20 & 0.13 & 0.03 & 0.08 & -0.03 & 0.07 & 0.06 & 0.03 & -0.08 & 0.03 \\
\footnotesize{NGC5822} & 0.06 & 0.15 & 0.05 & 0.01 & 0.29 & 0.30 & -0.04 & 0.01 & 0.02 & 0.22 & -0.04 & 0.02 & 0.68 & 0.72 & 0.10 &  & -0.06 & 0.04 & 0.21 & 0.19 & -0.10 & 0.03 \\
\footnotesize{NGC6475} & -0.05 &  & 0.04 &  & 0.21 &  & 0.03 &  & -0.14 &  &  &  & 0.04 &  &  &  & -0.11 &  & 0.14 &  & -0.04 &  \\
\footnotesize{NGC6494} & -0.03 & 0.02 & 0.04 & 0.02 & 0.10 & 0.04 & -0.02 & 0.01 & -0.12 & 0.02 & -0.03 & 0.02 & 0.26 & 0.02 & 0.08 & 0.06 & 0.06 & 0.03 & 0.09 & 0.02 & -0.07 & 0.01 \\
\footnotesize{NGC6633} & -0.12 & 0.04 & 0.04 & 0.01 & 0.14 & 0.02 & -0.03 & 0.01 & -0.10 & 0.01 & -0.02 & 0.01 & 0.19 & 0.03 & 0.12 & 0.03 & -0.02 & 0.01 & 0.12 & 0.02 & -0.08 & 0.01 \\
\footnotesize{NGC6705} & 0.05 & 0.09 & 0.03 & 0.02 & 0.04 & 0.07 & -0.00 & 0.02 & -0.04 & 0.08 & -0.02 & 0.03 & 0.12 & 0.05 & 0.02 & 0.06 & 0.19 & 0.05 & -0.04 & 0.04 & -0.01 & 0.02 \\
\footnotesize{NGC6940} & -0.17 & 0.11 & 0.02 & 0.01 & 0.06 & 0.00 & -0.02 & 0.01 & -0.14 & 0.02 & -0.02 & 0.03 & 0.07 & 0.03 & 0.11 & 0.06 & 0.04 & 0.13 & 0.08 & 0.01 & -0.05 & 0.01 \\
\footnotesize{NGC752} & -0.02 & 0.06 & 0.04 & 0.01 & 0.10 & 0.01 & -0.03 & 0.00 & -0.13 & 0.04 & -0.00 & 0.02 & 0.16 & 0.04 & 0.09 & 0.05 & -0.09 & 0.02 & 0.11 & 0.02 & -0.08 & 0.01 \\
\end{longtable}
\tablefoot{Empty values represent lack of enough suitable absorption lines and/or stars to compute the statistic.}
\end{landscape}
}% End onllongtab

\onllongtab{
\begin{landscape}
\footnotesize  % Switch from 12pt to 11pt; otherwise, table won't fit
\begin{longtable}{l|cc|cc|cc|cc|cc|cc|cc|cc|cc|cc}
\caption{Differential abundances with respect to M67 for the rest of elements not included in 'Msel'} \label{tab:abundances_summary_rest2}\\
 & \multicolumn{2}{c|}{[O/Fe]} & \multicolumn{2}{c|}{[Pr/Fe]} & \multicolumn{2}{c|}{[S/Fe]} & \multicolumn{2}{c|}{[Si/Fe]} & \multicolumn{2}{c|}{[Sm/Fe]} & \multicolumn{2}{c|}{[Sr/Fe]} & \multicolumn{2}{c|}{[V/Fe]} & \multicolumn{2}{c|}{[Y/Fe]} & \multicolumn{2}{c|}{[Zn/Fe]} & \multicolumn{2}{c}{[Zr/Fe]} \\
 & $\bar{x}$    &   $\sigma$  & $\bar{x}$    &   $\sigma$  & $\bar{x}$    &   $\sigma$  & $\bar{x}$    &   $\sigma$  & $\bar{x}$    &   $\sigma$  & $\bar{x}$    &   $\sigma$  & $\bar{x}$    &   $\sigma$  & $\bar{x}$    &   $\sigma$  & $\bar{x}$    &   $\sigma$  & $\bar{x}$    &   $\sigma$  \\
\hline
\footnotesize{Collinder350} &  &  & 0.11 &  &  &  & -0.07 &  &  &  & 0.26 &  & -0.09 &  & 0.18 &  & -0.22 &  & 0.13 &  \\
\footnotesize{IC2714} & -0.45 & 0.02 & -0.08 & 0.12 &  &  & -0.07 & 0.03 &  &  & 0.22 & 0.06 & -0.07 & 0.02 & 0.11 & 0.05 & -0.19 & 0.04 & 0.14 & 0.05 \\
\footnotesize{IC4651} &  &  & -0.09 & 0.07 & -0.08 & 0.08 & -0.01 & 0.02 & -0.12 & 0.09 & 0.01 & 0.02 & -0.03 & 0.03 & 0.03 & 0.04 & 0.01 & 0.07 & 0.11 & 0.09 \\
\footnotesize{IC4756} &  &  & 0.08 &  & 0.06 &  & -0.03 & 0.06 & -0.20 &  & 0.21 &  & -0.07 & 0.05 & 0.13 & 0.03 & -0.14 & 0.13 & 0.27 & 0.12 \\
\footnotesize{M67} & -0.25 & 0.18 & -0.06 & 0.05 & 0.02 & 0.05 & -0.00 & 0.01 & -0.12 & 0.10 & -0.03 & 0.04 & 0.00 & 0.03 & 0.01 & 0.05 & -0.01 & 0.03 & 0.10 & 0.07 \\
\footnotesize{Melotte111} &  &  &  &  & 0.01 & 0.03 & -0.02 & 0.01 & -0.06 & 0.06 &  &  & 0.03 & 0.08 & 0.09 & 0.04 & 0.05 & 0.08 & 0.18 & 0.06 \\
\footnotesize{Melotte20} &  &  &  &  & 0.06 &  & -0.03 &  &  &  &  &  & -0.02 &  & 0.07 &  & -0.13 &  &  &  \\
\footnotesize{Melotte22} &  &  &  &  & -0.01 & 0.06 & -0.02 & 0.01 &  &  &  &  & -0.02 & 0.01 & 0.14 & 0.06 & -0.10 & 0.17 &  &  \\
\footnotesize{Melotte71} &  &  & 0.10 & 0.14 &  &  & -0.08 & 0.05 & 0.44 & 0.64 & 0.20 &  & -0.14 & 0.14 & -0.01 & 0.13 & -0.10 & 0.13 & 0.15 & 0.01 \\
\footnotesize{NGC1817} &  &  & 0.09 & 0.11 &  &  & -0.08 & 0.02 &  &  & 0.20 & 0.08 & -0.09 & 0.02 & 0.04 & 0.03 & -0.17 & 0.05 & 0.23 & 0.19 \\
\footnotesize{NGC2099} & -0.65 &  & 0.04 & 0.01 &  &  & -0.10 & 0.03 & 0.43 &  & 0.15 & 0.23 & -0.12 & 0.04 & 0.14 & 0.10 & -0.13 & 0.11 & 0.06 & 0.06 \\
\footnotesize{NGC2251} & -0.47 & 0.06 & 0.07 & 0.01 &  &  & -0.06 & 0.03 & -0.03 &  & 0.30 &  & -0.08 & 0.04 & 0.10 & 0.04 & -0.23 & 0.01 & 0.13 & 0.01 \\
\footnotesize{NGC2287} & -0.39 & 0.03 & -0.07 & 0.01 &  &  & -0.03 & 0.01 & -0.17 & 0.01 & 0.22 &  & -0.04 & 0.01 & 0.12 & 0.02 & -0.13 &  & 0.15 & 0.01 \\
\footnotesize{NGC2360} & -0.43 & 0.04 & -0.08 & 0.03 &  &  & -0.07 & 0.01 &  &  & 0.16 & 0.03 & -0.08 & 0.01 & 0.08 & 0.03 & -0.21 & 0.01 & 0.12 & 0.01 \\
\footnotesize{NGC2423} & -0.39 & 0.04 & -0.00 & 0.08 &  &  & -0.04 & 0.03 &  &  & 0.13 & 0.03 & -0.06 & 0.04 & 0.11 & 0.03 & -0.07 & 0.08 & 0.08 & 0.01 \\
\footnotesize{NGC2447} & -0.38 & 0.07 & 0.15 & 0.01 & 0.04 &  & -0.02 & 0.04 & 0.03 & 0.08 & 0.32 & 0.01 & -0.06 & 0.05 & 0.13 & 0.04 & -0.12 & 0.11 & 0.17 & 0.05 \\
\footnotesize{NGC2477} & -0.62 &  &  &  &  &  & -0.10 &  &  &  &  &  & -0.13 &  & 0.16 &  & -0.28 &  & 0.23 &  \\
\footnotesize{NGC2539} & -0.46 & 0.03 & -0.08 & 0.08 &  &  & -0.07 & 0.01 & -0.11 & 0.04 & 0.17 & 0.03 & -0.07 & 0.02 & 0.12 & 0.05 & -0.17 & 0.02 & 0.13 & 0.03 \\
\footnotesize{NGC2547} &  &  &  &  &  &  & -0.04 &  &  &  &  &  & 0.13 &  & 0.24 &  & -0.26 &  &  &  \\
\footnotesize{NGC2548} & -0.52 & 0.12 & -0.07 & 0.16 &  &  & -0.08 & 0.05 &  &  & 0.16 &  & -0.12 & 0.04 & 0.07 & 0.05 & -0.11 & 0.05 & 0.11 & 0.01 \\
\footnotesize{NGC2567} & -0.42 & 0.05 & -0.04 & 0.12 &  &  & -0.09 & 0.02 & -0.19 &  & 0.26 & 0.07 & -0.07 & 0.03 & 0.15 & 0.08 & -0.25 & 0.05 & 0.18 & 0.05 \\
\footnotesize{NGC2632} & -0.42 & 0.02 & -0.12 & 0.04 & 0.00 & 0.08 & -0.01 & 0.02 & -0.05 & 0.14 & 0.05 & 0.02 & -0.04 & 0.04 & 0.04 & 0.06 & -0.06 & 0.14 & 0.13 & 0.09 \\
\footnotesize{NGC3114} & -0.41 & 0.04 & 0.01 & 0.10 &  &  & -0.04 & 0.04 & -0.03 & 0.01 & 0.23 & 0.06 & -0.07 & 0.03 & 0.12 & 0.03 & -0.14 & 0.12 & 0.12 & 0.03 \\
\footnotesize{NGC3532} & -0.51 & 0.07 & 0.05 & 0.16 &  &  & -0.09 & 0.05 & 0.11 & 0.01 & 0.24 & 0.05 & -0.06 & 0.03 & 0.13 & 0.06 & -0.22 & 0.07 & 0.17 & 0.02 \\
\footnotesize{NGC3680} & -0.36 & 0.06 & 0.10 & 0.02 & 0.01 & 0.07 & -0.03 & 0.02 & -0.08 & 0.10 & 0.13 & 0.03 & -0.01 & 0.05 & 0.07 & 0.03 & -0.06 & 0.02 & 0.10 & 0.02 \\
\footnotesize{NGC4349} & -0.37 & 0.04 & -0.21 & 0.21 &  &  & -0.07 & 0.05 & -0.15 & 0.14 & 0.15 & 0.15 & -0.04 & 0.04 & 0.04 & 0.07 & -0.24 & 0.14 & 0.10 & 0.07 \\
\footnotesize{NGC5822} &  &  & -0.24 & 0.13 &  &  & -0.09 & 0.02 &  &  & 0.57 & 0.63 & -0.09 & 0.04 & 0.39 & 0.43 & -0.18 & 0.02 & 0.40 & 0.38 \\
\footnotesize{NGC6475} &  &  &  &  & -0.08 &  & -0.02 &  & 0.03 &  &  &  & -0.01 &  & 0.08 &  & -0.01 &  & 0.07 &  \\
\footnotesize{NGC6494} &  &  & -0.22 & 0.12 &  &  & -0.06 & 0.02 & -0.22 &  & 0.18 & 0.02 & -0.07 & 0.01 & 0.06 & 0.01 & -0.20 & 0.04 & 0.12 & 0.01 \\
\footnotesize{NGC6633} & -0.43 &  & 0.04 & 0.10 &  &  & -0.06 & 0.01 &  &  & 0.23 & 0.03 & -0.08 & 0.01 & 0.11 & 0.03 & -0.19 & 0.02 & 0.15 & 0.02 \\
\footnotesize{NGC6705} & -0.47 & 0.06 & -0.23 & 0.14 &  &  & 0.01 & 0.04 & -0.04 & 0.11 & 0.02 & 0.06 & 0.01 & 0.04 & 0.13 & 0.05 & 0.17 & 0.11 & 0.06 & 0.03 \\
\footnotesize{NGC6940} & -0.45 & 0.03 & -0.02 & 0.01 &  &  & -0.06 & 0.00 & 0.04 & 0.06 & 0.15 & 0.03 & -0.08 & 0.01 & 0.15 & 0.01 & -0.12 & 0.02 & 0.10 & 0.03 \\
\footnotesize{NGC752} & -0.41 & 0.04 & 0.07 & 0.02 &  &  & -0.06 & 0.01 & -0.07 &  & 0.13 & 0.03 & -0.06 & 0.01 & 0.12 & 0.02 & -0.16 & 0.02 & 0.12 & 0.02 \\
\end{longtable}
\tablefoot{Empty values represent lack of enough suitable absorption lines and/or stars to compute the statistic.}
\end{landscape}
}% End onllongtab

\clearpage

\begin{appendix}

\section{Methods used in this paper}
\label{AppendixMethods}

A general presentation of several unsupervised classification methods that has been used in extragalactic astronomy can be found in \citet{Fraix-Burnet2015}. We present in this section some excerpts regarding the three methods used in this paper. 

There are two main categories of phylogenetic methods: the distance-based and the character-based. The ``characters'' are traits, descriptors, observables, parameters, variables or properties, which can be assigned at least two states characterizing the evolutionary stage of the objects for that character. For continuous variables, these states can be obtained through discretization. The most popular method for character-based approaches is cladistics (or Maximum Parsimony) and for the distance-based ones it is Neighbour Joining.

A detailed explanation of Maximum Parsimony can be found in \citet{Fraix-BurnetHouches2016} and a detailed illustration of its application on stellar evolutionary tracks is presented in an unpublished paper (Fraix-Burnet \& Thuillard 2014\footnote{\url{https://hal.archives-ouvertes.fr/hal-01703341}}).

\subsection{Cladistics or Maximum Parsimony}
\label{AppendixMethodsMP}

Cladistics when applied to domain outside of biology, like in astrocladistics, refers more generally to the classification of objects by a rooted or an unrooted tree. In that case, the tree represents possible relationships between objects (or classes of objects). Cladistics has been associated in the 80's to the search of a maximum parsimony tree.  Maximum Parsimony is a powerful approach to find tree-like arrangements of objects. The drawback is that the analysis must consider all possible trees before selecting the most parsimonious one. The computation complexity depends on the number of objects and character states, so that too large samples (say more than a few thousands) cannot be analyzed. 

In the standard approach to parsimony, the score $s_p$ of a tree corresponds, after labeling of the internal nodes, to the minimum number of edges $(u,v)$ with $c(u)\neq c(v)$, $c(u)$ being the character state at node $u$. The tree with the minimum score is searched for with some heuristics \citep{Felsenstein1984}. The maximum parsimony approach can be directly extended to continuous characters or values. To each internal node is associated a real value $f(u)$. The score s of a tree equals the sum over all edges of the absolute difference between those values:

\begin{equation}
 \label{eq:MPs}
    s = \sum_{e=(u,v) \epsilon E}  \lvert f(u) - f(v) \rvert
 \end{equation}

The success of a cladistics analysis much depends on the behavior of the input variables. In particular, it is sensitive to redundancies, incompatibilities, too much variability (reversals), and parallel and convergent evolutions. It is thus a very good tool for investigating whether a given set of input variables can lead to a robust and pertinent diversification scenario.

We wish to point out that Maximum Parsimony is based on the variables, not on pairwise distances like hierarchical clustering techniques or other phylogenetic approaches such as Neighbour Joining described in Appendix~\ref{AppendixMethodsNJ}.

\subsection{Neighbour Joining}
\label{AppendixMethodsNJ}

Among distance-based approaches, Neighbor-Joining is the most popular approach to construct a phylogenetic tree.
The Neighbor Joining Tree Estimation \citep[NJ,][]{NJ1987,NJ2006} is based on a distance (or dissimilarity) matrix. In this paper, we have taken the euclidean distance to compute this matrix. This method is a bottom-up hierarchical clustering methods. It starts from a star tree (unresolved tree). A ``corrected'' distance $Q(i,j)$ between objects $i$ and $j$ from the data set of $n$ objects, is computed from the distances $d(i,j)$:

\begin{equation}
\label{eq:NJ}
 Q(i,j) =  (n-2)d(i,j) - \sum_{k=1}^{n}d(i,k)  - \sum_{k=1}^{n}d(j,k) 
\end{equation}

The branches of the two objects with the lowest $Q(i,j)$ are linked together by a new node $u$ on the tree. This node replaces the pair $(i,j)$ in the subsequent iterations through the distance to any other object $k$: 

\begin{equation}
\label{eq:NJnode}
   d(u,k) = \frac{1}{2}\left[d(i,k)-d(i,u)\right] +  \frac{1}{2}\left[d(j,k)-d(j,u)\right]
\end{equation}

Neighbor-Joining minimizes a tree length, according to a criteria that can be viewed as a Balanced Minimum Evolution \citep{NJ2006}. For a tree metrics, Neighbor-Joining furnishes a simple algorithm to reconstruct a tree from the distance matrix. There is a large literature on how to best approximate a metrics by a tree metrics \citep[see for instance][]{Fakcharoenphol2003}.

\subsection{K-means}
\label{AppendixMethodsKM}

The k-means algorithm \citep{kmeans1967,kmeans2010} is not a phylogenetic tool, it is a partitioning approach that is simple and can be very efficient in some cases. 

The algorithm starts with $k$ centroids, $k$ corresponding to the number of clusters given a priori. It then assigns each data point to the closest (as measured by an Euclidean distance measure) centroid and when the clusters are built, the new $k$ centroids are computed and the process iterates until convergence. The result depends very much on the initial centroids, so that repeating the analysis with several initial choices (1000 in this paper) is thus necessary. However, consistency is not guaranteed if the data do not contain distinguishable and roughly spherical clusters. Some strategies have been devised to guess the best initial choice for the centroids \citep[e.g.][]{Sugar2003,Tajunisha2010} and many indices are available in the package $NbClust$ \citep{NbClust} of R \citep{R}.

\clearpage

\section{Contingency tables}
\label{TabContingencies}

In this Appendix we present the contingency tables for the four biggest open clusters (M67, NGC6705, IC4651 and NGC2632, see Table~\ref{TabClusters}) corresponding to the five clustering analyses performed in this paper: Sfull, Mfull and Msel with MP  (Sect.\ref{cladresult}), Msel with NJ (Sect.\ref{njresult}) and k-means (Sect.~\ref{kmresult}). 

Each tables provides the number of stars of each open cluster which are members of the groups given in the left column. The group index is arbitrary, ordered according to the best correspondence with the four open clusters and to their total number of stars for these clusters.

The precision and recall (sensitivity) of the classification is also computed in the tables for the four main open clusters. The precision gives the proportion of stars in a given group that belong to the same open cluster. The latter is supposed here to be the one that the largest number of stars in the given group. The precision is computed using all the open clusters (not shown in the tables).  The recall gives the proportion of stars among a given open cluster that belong to the group having most of its members. 

\clearpage

    \begin{table}
       \caption[]{Contingency table for the MP analysis on Sfull (Sect.~\ref{cladresult}) for the four biggest open clusters. \textit{n} is an arbitrary number index for the subset of groups depicted as gray boxes in Fig.~\ref{FigTrees} that contain at least one star of these four open clusters. Stars that are outside these gray boxes and hence are at the end of individual branches are not reported in the table. The last line is the recall values for each open clusters, and the last column is the precision for each group, computed using all the open clusters. }
          \label{TabContSfull}
\begin{tabular}{rrrrr|r}
             \hline
\textit{n} & {\footnotesize M67} & {\footnotesize NGC6705} & {\footnotesize IC4651} & {\footnotesize NGC2632} & Prec.\\ 
  \hline
  1 &  13 &   0 &   1 &   1 & 0.87 \\ 
  2 &   0 &  20 &   5 &   5 & 0.67 \\ 
  3 &   0 &   0 &   2 &   1 & 0.33 \\              
  4 &   0 &   0 &   6 &  12 & 0.67 \\ 
%  5 &   0 &   0 &   1 &   0 & \\ 
%  6 &   0 &   0 &   1 &   0 & \\ 
%  7 &   0 &   0 &   1 &   0 & \\ 
%  8 &   0 &   0 &   1 &   0 & \\ 
%  9 &   1 &   0 &   0 &   0 & \\ 
%  0 &   1 &   0 &   0 &   0 & \\ 
% 10 &   1 &   0 &   0 &   0 & \\ 
% 11 &   1 &   0 &   0 &   0 & \\ 
% 12 &   1 &   0 &   0 &   0 & \\ 
% 13 &   1 &   0 &   0 &   0 & \\ 
% 14 &   1 &   0 &   0 &   0 & \\ 
% 15 &   1 &   0 &   0 &   0 & \\ 
% 16 &   1 &   0 &   0 &   0 & \\ 
% 17 &   1 &   0 &   0 &   0 & \\ 
% 18 &   1 &   0 &   0 &   0 & \\ 
% 19 &   1 &   0 &   0 &   0 & \\ 
% 20 &   1 &   0 &   0 &   0 & \\ 
% 21 &   0 &   0 &   1 &   0 & \\ 
% 22 &   1 &   0 &   0 &   0 & \\ 
% 23 &   0 &   0 &   1 &   0 & \\ 
% 24 &   1 &   0 &   0 &   0 & \\ 
% 25 &   1 &   0 &   0 &   0 & \\ 
% 26 &   1 &   0 &   0 &   0 & \\ 
% 27 &   1 &   0 &   0 &   0 & \\ 
% 28 &   1 &   0 &   0 &   0 & \\ 
% 29 &   1 &   0 &   0 &   0 & \\ 
% 30 &   1 &   0 &   0 &   0 & \\ 
% 31 &   1 &   0 &   0 &   0 & \\ 
  \hline
Recall & 0.37 & 1.00 & 0.30 & 0.63 & \\
\hline  
          \end{tabular}
    \end{table}

\begin{table}[ht]
       \caption[]{Same as Table~\ref{TabContSfull} for the MP analysis of Mfull (Sect.~\ref{cladresult} and Fig.~\ref{FigTrees}).}
          \label{TabContMfull}
\begin{tabular}{rrrrr|r}
  \hline
\textit{n} & {\footnotesize M67} & {\footnotesize NGC6705} & {\footnotesize IC4651} & {\footnotesize NGC2632} & Prec.\\ 
  \hline
  1 &  17 &   2 &   5 &   0 & 0.68 \\ 
  2 &   0 &  18 &   0 &   0 & 1.00 \\ 
  3 &   0 &   0 &   5 &   0 & 0.16 \\ 
  4 &   0 &   0 &   0 &  13 & 1.00 \\ 
  5 &   5 &   0 &   0 &   0 & \\ 
  6 &   0 &   0 &   3 &   0 & \\ 
  7 &   0 &   0 &   0 &   1 & \\ 
%  8 &   1 &   0 &   0 &   0 & \\ 
%  9 &   0 &   0 &   1 &   0 & \\ 
% 10 &   0 &   0 &   1 &   0 & \\ 
% 11 &   0 &   0 &   1 &   0 & \\ 
% 12 &   0 &   0 &   1 &   0 & \\ 
 8 &   0 &   0 &   1 &   5 & \\ 
% 14 &   0 &   0 &   1 &   0 & \\ 
   \hline
 Recall & 0.74 & 0.9 & 0.26 & 0.68 & \\
   \hline
\end{tabular}
\end{table}

\begin{table}[ht]
       \caption[]{Same as Table~\ref{TabContSfull} for the MP analysis of Msel (Sect.~\ref{cladresult} and Fig.~\ref{FigTrees}).}
          \label{TabContMsel}
\begin{tabular}{rrrrr|r}
  \hline
& {\footnotesize M67} & {\footnotesize NGC6705} & {\footnotesize IC4651} & {\footnotesize NGC2632} & Prec.\\ 
  \hline
  1 &  18 &   2 &   6 &   1 & 0.60 \\ 
  2 &   0 &  18 &   1 &   0 & 0.95 \\ 
  3 &   5 &   0 &   4 &   0 & 0.20 \\ 
  4 &   0 &   0 &   4 &  17 & 0.89 \\ 
  5 &   0 &   0 &   0 &   1 & \\ 
%  6 &   0 &   0 &   1 &   0 & \\ 
%  7 &   0 &   0 &   1 &   0 & \\ 
%  8 &   0 &   0 &   1 &   0 & \\ 
%  9 &   0 &   0 &   1 &   0 & \\ 
     \hline
   Recall & 0.78 & 0.9 & 0.32 & 0.89 & \\
   \hline
\end{tabular}
\end{table}

\begin{table}[ht]
       \caption[]{Same as Table~\ref{TabContSfull} for the NJ analysis of Msel (Sect.~\ref{njresult} and Fig.~\ref{FigTreeNJ}).}
          \label{TabContMselNJ}
\begin{tabular}{rrrrr|r}
  \hline
\textit{n} & {\footnotesize M67} & {\footnotesize NGC6705} & {\footnotesize IC4651} & {\footnotesize NGC2632} & Prec.\\ 
  \hline
  1 &  14 &   0 &   6 &   1 & 0.64 \\ 
  2 &   0 &  17 &   0 &   0 & 1.00 \\ 
  3 &   7 &   2 &   5 &   0 & 0.27 \\ 
  4 &   0 &   0 &   3 &  17 & 0.85 \\ 
  5 &   1 &   0 &   3 &   0 & \\ 
  6 &   0 &   1 &   1 &   0 & \\ 
  7 &   0 &   0 &   0 &   1 & \\ 
%  8 &   1 &   0 &   0 &   0 & \\ 
%  9 &   0 &   0 &   1 &   0 & \\ 
       \hline
     Recall & 0.61 & 0.85 & 0.32 & 0.89 & \\
   \hline
\end{tabular}
\end{table}

\begin{table}[ht]
       \caption[]{Same as Table~\ref{TabContSfull} for the k-means analysis of Msel (Sect.~\ref{kmresult} and Fig.~\ref{Fig:kmeans30}).  }
          \label{TabContMselkm}
\begin{tabular}{rrrrr|r}
  \hline
\textit{n} & {\footnotesize M67} & {\footnotesize NGC6705} & {\footnotesize IC4651} & {\footnotesize NGC2632} & Prec.\\ 
  \hline
  1 &  13 &   0 &   3 &   0 & 0.76 \\ 
  2 &   0 &   7 &   0 &   0 & 1.00 \\ 
  3 &   0 &   7 &   0 &   0 & 1.00 \\ 
  4 &   1 &   0 &   4 &   0 & 0.67 \\ 
  5 &   0 &   0 &   1 &   7 & 0.88 \\ 
  6 &   0 &   0 &   0 &   7 & 1.00 \\ 
  7 &   5 &   0 &   2 &   0 & \\ 
  8 &   0 &   0 &   2 &   3 & \\ 
  9 &   0 &   2 &   3 &   0 & \\ 
 10 &   1 &   0 &   3 &   1 & \\ 
 11 &   0 &   2 &   1 &   0 & \\ 
 12 &   1 &   1 &   0 &   0 & \\ 
 13 &   0 &   0 &   0 &   1 & \\ 
 14 &   1 &   0 &   0 &   0 & \\ 
 15 &   1 &   1 &   0 &   0 & \\ 
       \hline
     Recall & 0.56 & 0.35 & 0.21 & 0.36 & \\
   \hline
\end{tabular}
\end{table}

\end{appendix}

\end{document}